\documentstyle[sprocl]{article}
\def\lsim{<\kern-2.5ex\lower0.85ex\hbox{$\sim$}\ }
\def\rsim{>\kern-2.5ex\lower0.85ex\hbox{$\sim$}\ }    
\def\lambdabar{\hbox{$\lambda$\kern-0.52em\raise+0.45ex\hbox{--}\kern+0.2em}}

\begin{document}

\title{THE SPONTANEOUS BREAKDOWN OF THE VACUUM}
\author{A.C. MELISSINOS}
\address{Department of Physics, University of Rochester,\\
Rochester, NY 14627 USA\\
meliss@pas.rochester.edu}

\maketitle\abstracts{We discuss the spontaneous breakdown of the vacuum by 
a strong electromagnetic field as observed in SLAC experiment E-144 
\cite{Burke}. We show that the data follow the Schwinger 
non-perturbative result obtained for a static field.}

\section{Pair Production in a Strong em Field}

In his 1951 paper, J. Schwinger \cite{Schwinger}
 predicted that an intense static 
electric field will break down the vacuum to produce $e^+e^-$ pairs.  This 
occurs when the field-strength approaches the critical value $E_c = 
m^2c^3/e\hbar = 1.3 \times 10^{16}$ V/cm.  We argue that this effect has been 
observed in the scattering of high energy electrons from the focus of an 
intense laser pulse \cite{Burke}.  

The experimental conditions \cite{Burke}
 differ from the original Schwinger premise 
in two respects:

(a)  The field strength in the laboratory is only 
$E \sim 3 \times10^{10}$ V/cm 
but reaches near-critical value in the rest frame of the 46.6 GeV incident 
electrons ($E^* = 2\gamma E \simeq 1.8 \times 10^5 E$).

(b) The field in the laboratory is not static but a well-defined coherent 
wave field.  However Brezin and Itzykson \cite{Brezin}
 have shown that spontaneous 
$e^+e^-$ pair production will also occur in a time-dependent field.  They 
derive the probability for pair production in both the perturbative (low 
field, $E^* < E_c$) and non-perturbative ($E^* \rsim E_c$) regimes.

It is convenient to introduce the normalized vector potential of the field
$$\eta = \left[ {e |\langle A_\mu A^\mu\rangle |\over m^2}\right]^{1/2} = 
{e E_{\rm rms}\over \omega_0mc}\eqno(1)$$
where $\omega_0$ is the angular frequency of the field which is assumed to 
be monochromatic and sinusoidal.  The probabilities per unit time -- unit 
volume derived \cite{Brezin} are
$$\eta \ll 1 \qquad w \simeq {\alpha E^2\over 4\hbar} \left({eE\over 
2m\omega_0c}\right)^{4mc^2/\hbar\omega_0}\eqno(2)$$

$$\eta \gg 1 \qquad w \simeq {\alpha E^2\over \pi\hbar} {\rm exp} 
\left( -{\pi m^2c^3\over 
e\hbar E}\right)\eqno(3)$$
Eqs. (2,3) have an immediate interpretation in physical terms.  When $\eta 
\ll 1$ we are in the perturbative regime and $n = 2mc^2/\hbar\omega_0$ 
is the number of photons required to produce the pair.  Thus the 
probability is proportional to the $n^{th}$ power of the square of the 
normalized vector potential: ($\eta^{2n}$).  When $\eta \gg 1$ the 
probability depends on the electric field strength through the 
singular expression exp $(-\pi E_c/E)$.
 In the static case this behavior can be interpreted~\cite{Heisenberg,Greiner}
 as quantum-mechanical tunneling through a potential $V_0 \sim 2 mc^2$.

In the intermediate case, $\eta \sim 1$,
$$w = {\alpha E^2\over \pi\hbar} {1\over g(\eta) + {1\over 2} g^\prime 
(\eta)/\eta} {\rm exp} \left[ - {\pi m^2c^3\over e\hbar E} 
g(\eta)\right]\eqno(4)$$
where the function
$$g(\eta) = {4\over \pi} \int^1_0 dy \left[ {1-y^2\over 1+ 
y^2/\eta^2}\right]^{1/2}\eqno(5)$$
smoothly interpolates between the two regimes.  In the 
non-perturbative regime it is customary to introduce the 
dimensionless parameter
$$\Upsilon = {E\over E_c} = {e\hbar E\over m^2c^3}\eqno(6)$$
If an electron moves through the electric field with 4-momentum $p$ 
($\gamma = p_0/mc)$ then in the electron's rest frame the parameter 
$\Upsilon$ takes the value
$$\Upsilon = {\sqrt{\langle (F^{\mu \nu} p_\nu)^2 \rangle} \over mc^2}
{1 \over E_c} \eqno(6^\prime)$$
where $F^{\mu \nu}$ is the field tensor. Eq. (6$^\prime$) can also be used 
when a high energy photon of 4-momentum $k_\nu$ traverses the field. For 
head-on collisions we can write
$$\Upsilon = 2 \gamma {e \hbar E \over m^2 c^3} = \eta {2(c p_0)
(\hbar \omega_0)
 \over (mc^2)^2} \eqno(6^{\prime \prime})$$
where $\gamma$ was defined previously and $\omega_0$ is the frequency of the 
em field. A few comments are in order: (1) $\Upsilon$ is a relativistic 
invariant describing the interaction of a particle with the electric field 
(2) $\Upsilon$ is well defined for a static field (3) It is a measure of 
the cm energy in the collision of the incoming particle with one laser 
photon (in units of the electron mass) multiplied by the normalized 
potential $\eta$.

In the experiment reported in ref. [1] electron-positron pairs were 
produced when 46.6 GeV/c electrons crossed the focus of a laser pulse 
of wavelength $\lambda = 527$~nm.  This observation can be interpreted as a 
two-step process in which first a photon backscatters off an electron to 
become a high energy $\gamma$-ray ($\omega_\gamma \sim 29$~GeV) 
and subsequently the $\gamma$-ray scatters from at least four laser 
photons to produce the pair.  The photon density in the focus is 
adequately high ($n_\omega \sim 2.5 \times 10^{26}\  {\rm cm}^{-3}$) so that 
multiphoton processes up to $n=5$ could be observed over the course of 
the experiment.  Support for this interpretation comes from plotting 
the positron yield as a function of $\eta$ and observing that it 
varies as $\eta^{2n}$ with $n = 5.1 \pm 0.2$ as expected from Eq. (2) 
if one replaces $\omega_0$ by $2\gamma\omega_0$.  In fact the data 
agree with an exact calculation of the multiphoton Breit-Wheeler 
equation \cite{Burke} as shown in Fig. 1.

To examine the alternative interpretation in terms of the spontaneous 
breakdown of the vacuum we wish to test Eq. (3).  Note that in the 
experiment, $\eta \sim 0.3$ namely one is between the two regimes.  We 
plot the data as a function of $1/\Upsilon$ as shown in Fig. 2 
where we use the form of Eq. (6$^\prime)$ for $\Upsilon$;
here we have also included data obtained at 49.1 GeV.
A fit to the $\Upsilon$ 
dependence of Eq. (4) yields for the factor in the exponent
$$\pi g (\eta) = 2.01 \pm 0.12 \pm 0.4\eqno(7)$$
the first error being statistical and the second systematic.  The 
prediction of Eqs. (4,5) for $\langle\eta\rangle = 0.25$ is $g(\eta) = 
0.58$ and thus $\pi g (\eta) = 1.82$.

However, the result of Eq. (7) must be corrected for two factors.  In Fig. 2 we 
used the rms value of the electric field to define $\Upsilon$ and 
$\eta$, whereas in ref. [3] the peak values are used.  Secondly the 
frame of reference in which the
high energy gamma and one photon collide should be used in 
 the definition of $\gamma$ entering Eq. (6$^\prime$);
namely $\Upsilon = \Upsilon_{\rm rms} \sqrt{2} (29.2/46.6)$.  We then 
find that $\Upsilon$ has to be reduced by a factor of 1.14, while 
$\eta$ must be increased by $\sqrt 2$.  This leads to
$$
 \pi g(\eta) = 1.77 \pm 0.35 \qquad {\rm observed} $$
$$\pi g(\eta) = 2.12 \qquad\qquad\quad {\rm predicted} \eqno(7^\prime)$$
We see that the dependence of the pair production rate on the field 
strength agrees with the predictions of refs. [1,3].

To estimate the positron yield predicted by Eq. (3) we must integrate the 
probability over volume and time.  
We associate a volume equal to $\lambdabar_c^3$ 
($\lambdabar_c$ 
is the Compton wavelength of the electron) for each electron 
crossing the focus and use $\Delta t = (\ell/c) (1/\gamma)$ for the 
time of interaction in the electron rest-frame; here $\ell$ is the 
length of the focus, $\ell = 2d/\sin\theta \sim 20 \;\mu$m. 
We then obtain

$$w = {\alpha \over \pi} {E^2_c \over \hbar} \Upsilon^2 
e^{-\pi g/\Upsilon} {1 \over g + {1 \over 2} g^\prime /\eta} \eqno(8)$$
with
$${\alpha \over \pi} {E^2_c \over \hbar} = 3.45 \times 10^{50}
\ {\rm cm}^{-3} {\rm s}^{-1} \eqno(8^\prime)$$
For $\Upsilon = 0.24$ and $\eta = 0.35$ we find for the probability per 
incident high energy $\gamma$-ray
$$\int w d^3x dt \simeq 4 \times 10^{-4} \eqno(9)$$
However per laser shot only $\sim 10^6$ $\gamma$-rays cross the laser focus 
and we must account for the fraction of $\gamma$-rays of sufficient energy 
to produce a pair ($\sim 10^{-2}$ of the total spectrum). These qualitative 
arguments predict a pair production rate of $\sim$ 4/laser shot as compared 
to the observed rate of 0.1/laser shot.

We make two additional remarks.  First, that eventhough the electric 
field seen in the electron rest-frame is time-dependent, the period 
is longer than the formation time of the pair by a factor of fifteen.  
In the rest-frame
$$T^* = {1\over\gamma} T = {1\over\gamma} {\lambda\over c} \sim 2 
\times 10^{-20} \; {\rm s}$$
whereas the quantum-mechanical uncertainty time associated with an 
energy fluctuation of the order of an electron mass is
$$\Delta t = {\lambdabar\over c} \sim 1.3 \times 10^{-21}\; {\rm s}$$
Thus one can treat the fields seen in the electron rest-frame as 
static as considered in ref. [2].  Note however that
this assumption is not needed in deriving Eqs. (2,3).

In the experiment of ref. [1] the energy of the electron and positron 
in the pair is provided by the incident high energy $\gamma$-ray. The 
presence of an incident particle resolves the issue of energy-momentum 
balance since it is known that a plane wave (for which $E^2 - B^2 = 0$) 
cannot produce pairs. On the other hand in a focused wave, there are 
regions near the focus \cite{Richards} where $E^2 - B^2 > 0$. The 
value of the invariant is approximately ${1 \over 2} (E/f^{\#})^2$ where
 $f^{\#}$ is the $f$-number of the focussing optics.

It would be of considerable interest to observe the breakdown of the vacuum 
without the participation of an incident particle. This would require an 
intensity of the em flux $I = 5 \times 10^{29}\  {\rm W/cm}^2$ to reach 
$\Upsilon = 1$.  Some future FEL's are planned to operate in the 1 ${\rm \AA
}$ region and one could consider extremely tight focussing (say to 10 ${\rm \AA
}^2$) to reach high intensity.  The peak power will be at best $P = 10^{11}
$ W so that $I = 10^{26}$ W/cm$^2$ which is still short of producing 
critical field in the laboratory frame.

\section{Breakdown of the Vacuum and the Fine Structure Constant}

The condition for the breakdown of the vacuum is that the electric 
field $E_c$ be such that an electron gains energy equal to its rest 
mass in one Compton wavelength
$$eE_c\lambdabar_c = mc^2\eqno(10)$$
Eq. (10) leads to the definition of the critical field $E_c$ introduced 
previously
$$E_c = {m^2c^3\over e\hbar}\eqno(11)$$
The above is a  statement on the interaction of the electron with
the field.  However for the pair to be produced there 
must also be sufficient energy in the field in a volume of order $V = 
\lambdabar_c^3$.  We therefore obtain a necessary condition on the
field energy 
$${1\over 2} \epsilon_0E^2_c\lambdabar_c^3 > 2mc^2\eqno(12)$$
Combining (9) and (10) leads to an upper limit on the value of the 
fine structure constant
$$\alpha = {e^2\over (4\pi\epsilon_0)\hbar c} < 
{1\over 16\pi}\eqno(13)$$
This inequality is satisfied in nature and appears to be
a necessary condition for the 
spontaneous breakdown of the vacuum.

\section*{Acknowledgments}

I thank Prof. E. M. Pachos for introducing me to the work of
 ref. [3]
and Prof. R. Sorkin for a clarifying discussion on Eq. (13);
 I thank Dr. Thomas Koffas for several calculations and the authors of 
ref. [1] for the use of the data presented in Figs. 1 and 2.

\vfill\eject

\ \ \

\vskip 5.5 truein

\noindent Fig. 1  Dependence of the positron rate on the 
laser field-strength parameter $\eta$. The rate is normalized to 
the number of Compton scatters inferred from the EC37 monitor.  The 
solid line is the prediction based on the numerical integration of the 
two-step process of laser backscattering followed by multiphoton 
Breit-Wheeler pair production.  From ref. [1], 46.6 GeV data.

\vfil\eject
\  \

\vskip 5.5 truein

\noindent Fig. 2 
  Number of positrons per laser shot as a function of 
${1/\Upsilon}$.  The solid line is a fit to the data of the form
$R_{e^+e^-} \propto {\rm exp} (-a/\Upsilon)$ and yields $a = 2.01 \pm 
0.12 \pm 0.4$
circles are for the 46.6 GeV data, squares for the 49.1 GeV data.
From Th. Koffas
 \lq\lq Positron Production in Multiphoton Light by Light Scattering"
 Ph.D Dissertation University of Rochester (1998) to be 
published.

\end{document}